\documentclass[showpacs,showkeys,preprint]{revtex4}

\begin{document}

\title{The effect of a cutoff on pushed and bistable fronts of the reaction diffusion equation}
\author{R.\ D.\ Benguria, M.\ C.\ Depassier and V. Haikala}
\address{        Facultad de F\'\i sica\\
       Pontificia Universidad Cat\'olica de Chile\\ Casilla 306,
Santiago 22, Chile}

\begin{abstract}
We give an explicit formula for the change of speed of pushed and bistable fronts of the reaction diffusion equation when a small cutoff is applied at the unstable or metastable equilibrium point. The results are valid for arbitrary reaction terms and include the case of density dependent diffusion.
\end{abstract}


\maketitle
\section{Introduction}
The effect of a cutoff on the speed of reaction diffusion fronts has received much attention since it was observed by Brunet and Derrida \cite{brunet-derrida} that fluctuations in propagating fronts which arise as a result of the discreteness in the number N of propagating particles can be modeled by introducing a small cutoff $\epsilon$ on the reaction term in the deterministic reaction diffusion equation. The cutoff parameter $\epsilon$ is related to the number of particles  by $\epsilon = 1/N$. Representing the effect of a finite population by a cutoff in continuum models has  been employed in other contexts as well \cite{Cohenandothers} (and references therein).
 We refer to \cite{Panjareview} for a very complete review and references on this topic. The effect of fluctuations and noise is particularly important   when considering front propagation into a metastable state.  The speed of the front without a cutoff may be quite small, zero in fact at the Maxwell point, in which case either noise \cite{Engel} or a cutoff, as we see below,  induces front propagation. The effect of fluctuations in front propagation of reaction diffusion systems are of interest in a variety of systems,  we expect that the present results will be useful in determining the validity of representing  fluctuations by a simple cutoff in the reaction term.

The effect of a cutoff on the speed of pulled fronts has been studied extensively \cite{brunet-derrida,Panjareview,Panja1,Mendez}, less attention has been paid to the effect of a cutoff on pushed and bistable fronts. For pulled fronts, with or without cutoff, the speed can be calculated by a linear analysis at the edge of the front.    The effect of a cutoff on pushed and bistable fronts has been studied for the exactly solvable case of the Nagumo reaction term  $f(u) = u(1-u)(u-a)$ which, for different values of the parameter $a$, describes bistable, pushed and pulled fronts (this is also known as Schl\"ogl reaction term when written in the variable $\rho=2 u -1$). It was shown in \cite{Mendez,kns} that the shift in the speed has a power law dependence on the cutoff parameter $\epsilon$ in contrast to the inverse square logarithmic dependence on the cutoff parameter for pulled fronts found by Brunet and Derrida. It has also been shown using a variational approach \cite{Mendez} that a cutoff slows down pulled and pushed fronts but speeds up bistable fronts.
The purpose of this work is to provide an explicit expression for the shift of the speed of pushed and bistable fronts with a cutoff for arbitrary reaction terms including the case of density dependent diffusion.   We show  that the shift in the speed is given by
$$
\delta c = - K f'(0)\, \epsilon^{1+\lambda}
\qquad \mbox{where}\quad
\left\{ \begin{array}{ll}
              -1 < \lambda < 0
             & \mbox{for pushed fronts} \\
             \,\,\,\,\,0 <\lambda <1 &\mbox{for bistable fronts}
           \end{array}
\right.
$$
where the constants $K$ and $\lambda$ are  independent of $\epsilon$. When the slope of the reaction term vanishes at $u=0$ we find
$$
\delta c = -K f(\epsilon) \qquad \mbox{when $f'(0)=0$.}
$$
We give explicit expressions for the constants $\lambda$ and $K$ which depend only on the front without a cutoff. The shift in the speed with density dependent diffusion is contained in this last case, $f'(0)=0$, as we indicate below.

Before proceeding with the actual derivation we need to recall some known results on the speed of fronts of the reaction diffusion equation.
We consider the reaction diffusion equation
\[u_t = u_{xx} + f(u) \qquad \mbox{with}\qquad f(0)=f(1)=0,
\]
where the reaction term  $f(u)$ satisfies additional conditions depending on the physical problem under consideration. We shall consider two generic classes.
The first, type I, satisfies $f>0$ in $(0,1)$. To this category belong pulled and pushed fronts. Type II,  bistable reaction terms, satisfy $f(u) <0$ for $u$ in $(0,a)$, $f>0$ in $(a,1)$ with $\int_0^1 f(u) du >0.$

For both types of reaction terms, sufficiently localized initial conditions evolve into a monotonic front \cite{AW78}. For reaction terms of type I the system evolves  into the monotonic front of minimal speed. If this minimal speed is that obtained from the linear analysis at the edge of the front \cite{KPP}, that is, $c_{min}\equiv c_{KPP} = 2 \sqrt{f'(0)}$, the front is called pulled. If this minimal speed is greater than $c_{KPP}$ the front is called pushed. For reaction terms
of type II there is a unique speed for which a monotonic front exists. It has been shown \cite{BD96c} that the asymptotic speed of pushed and bistable fronts  is given by the variational formula
\begin{equation}
c^2\,=  \max \left( 2\, \frac{\int_0^1  f\, g\, du}{\int_0^1
 (-g^2/g') du} \right),
\label{var}
\end{equation}
where the maximum is taken over all positive decreasing
functions $g$
in $(0,1)$ for which the integrals exist. The maximum is attained for a  trial fuction $g= \hat g$ (unique, up to a multiplicative constant) which, close to $u=0$, diverges as
\begin{equation}
\hat g \approx \frac{1}{u^{c/m}},
\label{ghat}
\end{equation}
where
$$
m = \frac{1}{2} \left[ c + \sqrt{c^2 - 4 f'(0)} \right].
$$
For pulled fronts the maximum in (\ref{var}) is not attained and the speed is given instead by the supremum of  $2\,{\int_0^1  f\, g\, du}/{\int_0^1
 (-g^2/g') du}$ over the class of functions mentioned above.

For pushed and bistable fronts, the existence of a variational principle allows one to use the Feynman--Hellman theorem to calculate the dependence
of the speed on parameters of the reaction term. We shall use this theorem to study the effect of a cutoff
on pushed and bistable fronts. Suppose that the reaction term $f$ depends on a parameter $\alpha$ (i.e., $f=f(u,\alpha)$). In the context of the variational principle (\ref{var}), the Feynman--Hellman theorem reads as follows,
\begin{equation}
\frac{\partial c^2}{\partial \alpha} = 2 \frac{\int_0^1  \frac{\partial f}{\partial \alpha}(u,\alpha)\, \hat g(u,\alpha)\, du}{\int_0^1
 (-\hat g^2/\hat g') du},
\label{fh}
\end{equation}
where $\hat g(u,\alpha)$ is the function (unique up to a multiplicative constant) that yields the maximum in (\ref{var}) at the given parameter $\alpha$. Notice that the Feynman--Hellman theorem holds only if the maximum is attained, which is not the case for pulled fronts.
In what follows we use the Feynman--Hellman theorem taking the cutoff $\epsilon$ as the parameter.

Consider a reaction term with a cutoff of the form $ f(u) \Theta(u-\epsilon)$, where the reaction term $f(u)$ without a cutoff gives rise to a pushed or bistable front (here, $\Theta(x)$ denotes the Heaviside step function). The Feynman-Hellman theorem tells us that
\begin{equation}
\frac{\partial c^2}{\partial \epsilon} = 2 \frac{\int_0^1  \frac{\partial f(u) \Theta(u-\epsilon)}{\partial \epsilon}\, \hat g(u,\epsilon)\, du}{\int_0^1
 (-\hat g^2/\hat g') du},
= - 2 \frac{f(\epsilon)\hat g(\epsilon,\epsilon)}{\int_0^1
 (-\hat g^2/\hat g') du}
\end{equation}
In the expression above $\hat g(\epsilon,\epsilon)$ is the optimizing function for the speed of the front with the reaction term $f(u)\Theta(u-\epsilon)$. We are interested in the speed when $\epsilon$ is small therefore,  in leading order, $\hat g(u,\epsilon) = \hat g(u,0)$. The function $\hat g(u,0)$ is the optimizing function for the reaction term $f(u)$ which we call simply  $\hat g_0(u)$.
To leading order in $\epsilon$ we obtain finally
$$
c \frac{d c}{d \epsilon}\biggm|_{\epsilon=0} = -\frac{f(\epsilon)\hat g_0(\epsilon)}{{\int_0^1
 (-\hat g_0^2(u)/\hat g_0'(u)) du}}.
 $$

 The shift in the speed is then
 $$
 \delta c = \epsilon  \frac{d c}{d \epsilon} \biggm|_{\epsilon=0} = - K \epsilon f(\epsilon)\hat g_0(\epsilon),
 $$
 where the proportionality constant K, which is independent of $\epsilon$, is given by
 $$
 K = \left[c_0 \int_0^1
 (-\hat g_0^2/\hat g_0') du\right]^{-1}= \frac{c_0}{2\, \int_0^1  f\, \hat g_0\, du}.
 $$
 In the formula above, $c_0$ is the speed of the front in the absence of cutoff. Using (\ref{ghat}) we have that
 $$
 \delta c = - K f(\epsilon) \epsilon^{1 - c/m}.
 $$
 Replacing the value of $c/m$,  we obtain finally,
 $$
 \delta c = \left\{ \begin{array}{ll}
              - K \, f(\epsilon)
             & \mbox{if}\qquad f'(0) =0 \\
             - K \, f'(0)\, \epsilon^{1+ \lambda} &\mbox{if}\qquad f'(0) \neq 0
           \end{array}
\right.,
$$
where
$$
\lambda = \frac{\sqrt{1 - 4 f'(0)/c_0^2} -1}{\sqrt{1 - 4 f'(0)/c_0^2} + 1}.
$$
In the expression above for $\delta c$ we used $f(\epsilon)\approx \epsilon f'(0)$ in leading order. For pushed fronts $f'(0) >0$ therefore $-1 <\lambda < 0$ and $\delta c <0$. For bistable fronts
$f'(0) < 0$ therefore $0< \lambda <1$ and $\delta c > 0$.

The precise value of the constants $K$ and $\lambda$ can be determined analytically when the solution
 for the speed in the absence of cutoff is known, that is when $\hat g_0$ is known

As an example we may apply the above result to the Nagumo reaction term
$$ f(u) = u (1-u) (u - a)
$$
for which an exact solution is known.
This is the case studied previously by other methods \cite{kns,Mendez,Panjareview}.

For $0<a<1/2$ this is a bistable reaction term. For negative values of $a$ this is a reaction term of type I which  for $-1/2<a<0$ gives rise to a pushed front.
The speed without the cutoff is given by
$$ c_0 = \frac{1}{ \sqrt{2}} - a \sqrt{2} $$
which is obtained from the variational principle (\ref{var}) with the trial function \cite{BD96c}
$$
\hat g_0(u) = \left( \frac{{1-u}}{ u} \right)^{1 -
2 a}.$$
For this reaction term $f'(0) = - a$. The value of $K$ is
$$
K = \left[c_0 \int_0^1 (-g_0^2/g_0') dq\right]^{-1} =
 \frac{\sqrt{2}\,\Gamma(4)}{\Gamma(1+ 2 a)\Gamma(3 - 2 a)}.
$$and $\lambda = 2 a$. Therefore,
$$
\delta c = \frac{\sqrt{2}\,\Gamma(4)}{\Gamma(1+ 2 a)\Gamma(3 - 2 a)}\,a \,\,\epsilon^{1 + 2 a}.
$$
The power dependence of the shift on $\epsilon$ is in agreement with previous results \cite{kns}, the magnitude has not been calculated elsewhere. Notice that close to the Maxwell point a=1/2, where the speed of the front without cut-off vanishes, the speed of the front is due to the cut-off alone.

When $a=0$, $f'(0)=0$ and, in leading order, the shift of the speed is given by
$$
\delta c = {3\sqrt{2}} \,\, \epsilon^2.
$$

As a second example consider the Fisher-Kolmogorov equation with density dependent diffusion
\begin{equation}
u_t= (D(u) u_x)_x + u (1-u),
\label{density}
\end{equation}
where the diffusion coefficient $D(u)$ satisfies $D(0)=0,\, D'(u)>0$. When the diffusion coefficient is not constant it is not possible to determine the speed of the front from linear considerations.  Moreover the wave of minimal speed is sharp, that is, it does not decay exponentially at infinity. One of the most studied cases is a power law dependence of the form $D(u) = u^m$. The exact solution is known for the value $m=1$, case for which the asymptotic speed of the front is given by $c=1$ \cite{Newman80,Aronson}. The speed of traveling fronts of equation (\ref{density}) is equal to the speed of the reaction diffusion equation with constant diffusion but with a reaction term $f(u)= D(u) u(1-u)$ \cite{BD96c} and it belongs to the case $f'(0)=0$. Therefore, a cutoff $\epsilon$ produces a shift in the speed given by
$$
\delta c = - K D(\epsilon)f(\epsilon)= -K \epsilon D(\epsilon),
$$
in leading order. Again, the constant $K$ is independent of $\epsilon$ and it can be determined if the exact solution without a cutoff is known.

In summary, we have established the effect of a cutoff on the speed of fronts of the reaction diffusion equation for all fronts which are not pulled in the absence of a cutoff. This has been done in a simple, unified way, making use of a variational principle for the asymptotic speed of the fronts. We find not only the dependence on the cutoff but an explicit expression for the shift. The method used to obtain these results is the Feynman--Hellman theorem, which enables one to determine the effect of varying any parameter of the reaction term. For pulled fronts, the speed is given by the supremum of an integral expression, not the maximum, hence the Feynman--Hellman theorem is not valid for them. For pulled fronts the effect of the cutoff can also be calculated from the variational expression but by directly solving the Euler-Lagrange equation in the linear approximation \cite{BDcutoff}. The approach used here to calculate the effect of the cutoff on the speed of pushed fronts can be used for the reaction convection diffusion equation and for the hyperbolic reaction diffusion equation, for which integral variational principles have been formulated \cite{BDhyp,BDconvec}.

\section*{Acknowledgements}
We acknowledge partial support of Fondecyt (CHILE) projects 106--0627 and 106--0651, and  CONICYT/PBCT Proyecto Anillo de Investigaci\'on en Ciencia y Tecnolog\'ia ACT30/2006.


\begin{thebibliography}{10}

\bibitem{brunet-derrida} E. Brunet and B. Derrida, Phys. Rev. E {\bf 56}, 2597
(1997).

\bibitem{Cohenandothers}
E. Cohen, D. A. Kessler, H. Levine, Phys. Rev. Lett. {\bf 94} 098102 (2005).


\bibitem{Panjareview}
D. Panja, Physics Reports {\bf 393}, 87 (2004).

\bibitem{Engel}
A. Engel, Phys. Lett. A {\bf 113}, 139 (1985).

\bibitem{Panja1} D. Panja and W. van Saarloos, Phys. Rev. E {\bf 65},
057202 (2002).

\bibitem{Mendez}
 V. Mendez, D. Campos and E. P. Zemskov,
 Phys. Rev. E {\bf 72}056113 (2005).

\bibitem{kns} D. A. Kessler, Z. Ner, and L. M. Sander,
Phys. Rev. E. {\bf 58}, 107 (1998).

\bibitem{AW78}
D.~G. Aronson and H. F. Weinberger,
\newblock { Adv. Math.} {\bf 30}, 33 (1978).

\bibitem{KPP}
A. Kolmogorov, I. Petrovsky, and N. Piskunov,
\newblock { Bull.  Univ. Moscow, Ser. Int. A} {\bf 1}, 1 (1937).


\bibitem{BD96c}
R.~D. Benguria and M.~C. Depassier, Phys. Rev. Lett. {\bf 77},
1171 (1996).

\bibitem{Newman80}
W. I. Newman, J. Theor. Biol. {\bf 85}, 325, (1980)

\bibitem{Aronson}
D. G. Aronson, in {\it Dynamics and Modelling of Reacting
Systems}, edited by W. Stewart et al. (Academic, New York,
1980).

\bibitem{BDcutoff}
R.~D. Benguria and M.~C. Depassier, preprint 2007.

\bibitem{BDhyp}
R.~D. Benguria and M.~C. Depassier, Phys. Rev. E {\bf 66}, 26607 (2002).

\bibitem{BDconvec}
R.~D. Benguria, M.~C. Depassier and V. M\'endez, Phys. Rev. E {\bf 69}, 031106 (2004).



\end{thebibliography}
\end{document}